\documentclass[twocolumn,preprintnumbers,amsmath,amssymb]{revtex4}

\usepackage{graphicx}
\usepackage{bm}

\begin{document}

\title{Two-dimensional imaging of the spin-orbit effective magnetic field}

\author{L. Meier}
\altaffiliation{also at Solid State Physics Laboratory, ETH
Zurich, 8093 Zurich, Switzerland}
\author{G. Salis} \email{gsa@zurich.ibm.com}
\affiliation{IBM Research, Zurich Research Laboratory,
S\"aumerstrasse 4, 8803 R\"uschlikon, Switzerland}
\author{E. Gini$^1$, I. Shorubalko$^2$, and K. Ensslin$^2$}
\affiliation{$^1$FIRST Center for Micro- and Nanosciences, ETH
Zurich,
8093 Zurich, Switzerland\\
$^2$Solid State Physics Laboratory, ETH Zurich, 8093 Zurich,
Switzerland}

\date{September 13, 2007}

\begin{abstract}
We report on spatially resolved measurements of the spin-orbit
effective magnetic field in a GaAs/InGaAs quantum-well. Biased
gate electrodes lead to an electric-field distribution in which
the quantum-well electrons move according to the local orientation
and magnitude of the electric field. This motion induces Rashba
and Dresselhaus effective magnetic fields. The projection of the
sum of these fields onto an external magnetic field is monitored
locally by measuring the electron spin-precession frequency using
time-resolved Faraday rotation. A comparison with simulations
shows good agreement with the experimental data.
\end{abstract}

\maketitle

\begin{figure}
\includegraphics[width=80mm]{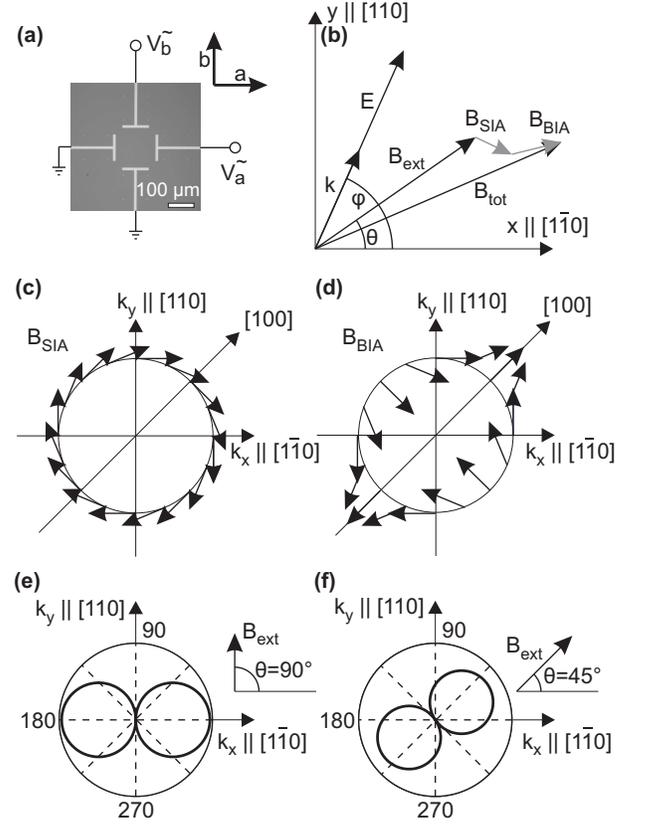}
\caption{\label{fig:fig1} (a) Microscope image of sample structure
and schematic gate connections. (b) Relevant fields and angles. A
magnetic field $\boldsymbol{B}_\textrm{ext}$ is applied at an
angle $\theta$ with respect to the $[1\overline{1}0]$ axis, and an
electric field $\boldsymbol{E}$ at an angle $\varphi$. (c,d)
Orientation of (c) Rashba and (d) Dresselhaus effective magnetic
fields for the electron wave vector $k$ on a unit circle. (e,f)
Polar plot of $A(\theta,\varphi)$ as a function of $\varphi$ for
$\theta = 90^\circ$ and $45^\circ$, as calculated for
$B_\textrm{SIA}$\,=\,-4\,mT and $B_\textrm{BIA}$\,=\,21\,mT
(sample 1). Depending on $\theta$, $A$ is proportional to either
the component of $\boldsymbol{E}_0$ perpendicular or approximately
parallel to $\boldsymbol{B}_\textrm{ext}$.}
\end{figure}

In the reference frame of a moving electron, electric fields
transform into magnetic fields, which interact with the electron
spin and couple it to the electron's orbital motion, leading to
spin-orbit (SO) interaction. In crystals lacking an inversion
center such as GaAs, effective magnetic fields due to bulk
inversion asymmetry (BIA) were predicted by
Dresselhaus~\cite{Dresselhaus1955}. In heterostructures, structure
inversion asymmetry (SIA) leads to an effective magnetic field
called Rashba term~\cite{Bychkov1984}. Both contributions have
been studied extensively (for a review, see
Ref.~\cite{Winklerbuch}) and are thought to play a crucial role in
future spintronic devices, because the coupling of the orbital and
the spin degrees of freedom opens a new way to spin manipulation,
for example by flipping spins with oscillating electric
fields~\cite{RashbaPRL2003, RashbaAPL2003,
Duckheim2006,Golovach2006}. The interplay between Dresselhaus and
Rashba SO interaction has been studied~\cite{Duckheim2007,
Bernevig2006} and proposed for use in a spin
transistor~\cite{Schliemann2003}. For finite electron wave
numbers, SO interaction leads to a spin splitting at zero external
magnetic field, which is observable as a beating of Shubnikov-de
Haas oscillations~\cite{Das1989,Luo1990,
Engels1997,Schapers1998,Hu1999,Pfeffer1999,Brosig1999} and can be
used to experimentally determine the strength of the total SO
interaction. The zero-field spin splitting also results in a
spin-selective momentum scattering and can lead to spin-dependent
photocurrents~\cite{Ganichev2004}, which allow the determination
of the ratio between the Rashba and Dresselhaus contributions by
studying their directional dependence. Since SO interaction is the
main reason for spin relaxation in GaAs quantum well (QW) samples,
the directional dependence of spin relaxation reveals the relative
strength of the Rashba and Dresselhaus terms~\cite{Averkiev2006}.
A more direct method to get access to the SO fields is to impose a
drift momentum on the conduction-band electrons by applying an
in-plane electric field. This leads to two different effects: In a
steady-state situation, the spins are oriented along the SO
field~\cite{Edelstein1990,Aronov1991,KatoPRL2004,Silov2004}. If
the spins of an ensemble of such drifting electrons are polarized
by means of an optical pump pulse, a coherent spin precession
about the SO field takes place, which has been observed using
time-resolved Faraday rotation (TRFR) in a bulk GaAs
epilayer~\cite{KatoNature2004}. Using TRFR, we have recently shown
that the absolute values of both the Rashba and the Dresselhaus
field can be determined in a two-dimensional electron gas by
studying the precession frequency of electron spins as a function
of the electron's direction of motion with respect to the crystal
lattice~\cite{MeierNatPhys2007}. An oscillating electric field was
applied at an arbitrary direction in the plane of an InGaAs
quantum well (QW) using two pairs of opposed electric gates
arranged perpendicularly to each other and enclosing a square area
of QW electrons [see Fig.~\ref{fig:fig1}(a)].

Here, we study the spatial distribution of the total SO effective
magnetic field between and outside the four gate electrodes that
are used to generate the in-plane electric field. We show that the
measured maps of spin-precession frequency can be well explained
by assuming that the electrons move along the spatially varying
electric field and that their spins perceive a SO effective
magnetic field given by their local drift momentum. Depending on
the orientation of an external magnetic field with respect to the
crystal axis, the precession frequency becomes sensitive to drift
momentum along a fixed in-plane direction, allowing to spatially
map the corresponding momentum component. A simulation with fixed
values for the Rashba and Dresselhaus coefficients as determined
in Ref.~\cite{MeierNatPhys2007} leads to good agreement with the
measured precession-frequency maps. Specifically, the results
confirm that the electron drift momentum in the center of the
gates can be described by the superposition of two perpendicular
components given by the two orthogonal gate biases, as it was
assumed in Ref.~\cite{MeierNatPhys2007}. Away from the center,
however, the electric field and thus the spin-precession frequency
start to vary. Large SO-induced effects on the spin-precession
frequency can still be detected far away from the region between
the gates, because of the connection lines to the gates. This
illustrates the importance of a precise definition of the
electrical connection scheme in a larger set-up and the difficulty
of applying electric fields only at given positions of a
two-dimensional electron gas.

The samples are either a 20- or 43-nm-wide InGaAs/GaAs QW as
described in Ref.~\cite{MeierNatPhys2007} (sample 1 and 2). We
experimentally determine $B_\textrm{tot} =
|\boldsymbol{B}_\textrm{tot}| = h\nu/g\mu_B$ by measuring the
electron spin-precession frequency $\nu$ of optically excited
conduction-band electrons using scanning TRFR ($h$ denotes
Planck's constant, $g$ the electron $g$-factor, and $\mu_B$ the
Bohr magneton). With a first, circularly polarized pump pulse
tuned to the absorption edge of the QW ($\lambda = 870$~nm,
average power $400\,\mu$W, pulse width of 2\,ps, and repetition
rate of 80\,MHz), we create a spin polarization in the QW
conduction band perpendicular to the QW plane. A pump-probe delay
time $\Delta\tau$ later, we probe the spin polarization with a
linearly polarized probe pulse (average power $60\,\mu$W) and
monitor the rotation angle $\theta_F$ of its polarization plane.
This Faraday rotation angle can be fit to $\theta_F = \theta_0
\exp{(-\Delta\tau/T_2^\star)}\cos{(2\pi\nu\Delta\tau)}$ to yield
both the spin-precession frequency $\nu$ and the spin-coherence
time $T_2^\star$. By spatially moving the sample relative to the
laser beams, the spin dynamics can be studied with a spatial
resolution limited by the diameter of the beam in the focus, here
approximately 15~$\mu$m. Scanning Kerr microscopy has been used in
Refs.~\cite{PRB.68.041307} for spatially resolving nuclear
imprinting effects and in Refs~\cite{Crooker2005b,Crooker2005} for
studying spin transport in GaAs epilayers.

On top of the sample, four gates and the corresponding connection
lines were defined using standard electron-beam lithography,
evaporation of a 10-nm-thick Ti adhesion layer and a 80-nm-thick
Au layer, and lift-off techniques. As shown in
Fig.~\ref{fig:fig1}{(a)}, opposite gate-pairs were connected to
phase-locked oscillators, one each for the two perpendicular
directions $a$ and $b$. Two neighboring gates were grounded, and
by adjusting the voltage amplitudes $V_a$ and $V_b$ on the other
two gates, an oscillating electric field $\boldsymbol{E}(t) =
\boldsymbol{E}_0\sin{(2\pi ft)}$, $f=160$~MHz, is induced in the
QW plane. In the center of the four electrodes, $\boldsymbol{E}_0$
points at an angle, with respect to the $a$-axis, given by
$\arctan V_b/V_a$. We have fabricated different samples in which
the $a$-axis is oriented along either $[1\overline{1}0]$ or
$[100]$ of the semiconductor crystal. We define a coordinate
system with $x \| [1\overline{1}0]$ and $y \| [110]$, and label
the angle of $\boldsymbol{E}_0$ with the $x$-axis as $\varphi$,
see Fig.~\ref{fig:fig1}(b). As long as $1/f$ is large compared to
the mean collision time of the electrons in the QW ($\approx
1$~ps), the electrons adiabatically follow $\boldsymbol{E}(t)$ and
their drift wave vector is given by $\boldsymbol{k} = m^\star \mu
\boldsymbol{E}(t)/hbar$, with $m^\star$ the effective electron
mass and $\mu$ the electron mobility. Furthermore, the small
collision time prevents the electrons from leaving the laser focus
during the time $1/f$. Because of their non-vanishing $k$-vector,
the electrons are subject to SO effective magnetic fields, which
in two-dimensional systems are given by~\cite{GanichevReview2003,
Winklerbuch}
\begin{equation}\label{eq:SIABIA}
   \boldsymbol{B}_{\textrm{SIA}} = \frac{2\alpha}{g \mu_B} \binom{k_y}{-k_x} \quad \textrm{and}
   \quad
   \boldsymbol{B}_{\textrm{BIA}} = \frac{2\beta}{g \mu_B} \binom{k_y}{k_x},
\end{equation}
with $\boldsymbol{k} = (k_x, k_y)$ and $\alpha$ and $\beta$ the
Rashba and Dresselhaus coupling constants, respectively. The
geometrical dependence on $\boldsymbol{k}$ is shown in
Fig.~\ref{fig:fig1}{(c) and (d)}. Whereas
$\boldsymbol{B}_{\textrm{SIA}}$ is always perpendicular to
$\boldsymbol{k}$, $\boldsymbol{B}_{\textrm{BIA}}$ points along
$\pm\boldsymbol{k}$ for the $[100]$ and $[010]$ directions. When
analyzing the spin precession, both SO contributions can be added
to an external magnetic field
$\boldsymbol{B}_\textrm{ext}$~\cite{Kalevich1990,EngelPRL2007},
which we apply in the plane of the QW and at an angle $\theta$
with the $x$-axis, as illustrated in Fig.~\ref{fig:fig1}{(b)}. For
all data presented, we choose $B_\textrm{ext} = 0.958$~T.

If $B_{\textrm{SIA}}$, $B_{\textrm{BIA}} \ll B_\textrm{ext}$, the
total magnetic field $B_\textrm{tot}$ can be expressed
by~\cite{MeierNatPhys2007}
\begin{equation}\label{eq:RootExpanded}
    B_\textrm{tot}(t) \approx B_\textrm{ext} + A(\theta,\varphi)\sin{(2\pi f t)}.
\end{equation}
Here, $t$ is the time delay between the electric-field oscillation
and the pump pulse (known up to an offset $t_0$, which is constant
in all experiments), and
\begin{equation}\label{eq:AB}
\begin{split}
  A(\theta,\varphi)  = \,&\left(B_\textrm{BIA} + B_\textrm{SIA}\right)\cos{\theta}\sin{\varphi}\\
  + &\left(B_\textrm{BIA} -
  B_\textrm{SIA}\right)\sin{\theta}\cos{\varphi}.
\end{split}
\end{equation}

\begin{figure}
\includegraphics[width=80mm]{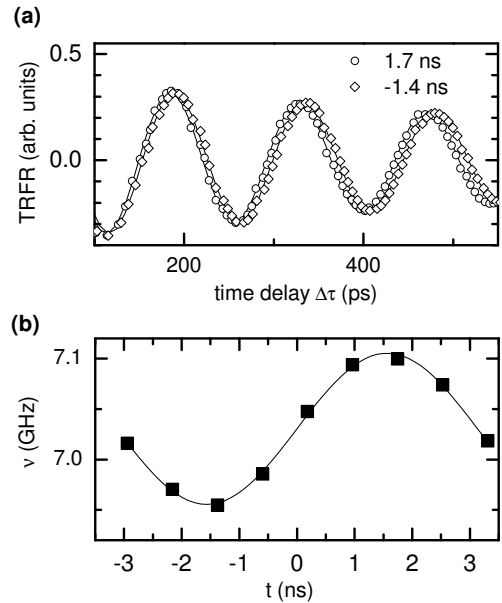}
\caption{\label{fig:fig2} (a) Measured TRFR vs. $\Delta \tau$ on
sample 1 with an electric field oscillating at 160\,MHz along the
$[100]$ axis, and $\theta$\,=\,90$^{\circ}$. Two curves are shown
for two different times $t$ at which the electric field has
opposite signs (open symbols). The solid lines are a fit to an
exponentially decaying oscillation. In (b), the obtained
precession frequency $\nu$ is plotted vs $t$ (symbols). The solid
line represents a fitted harmonic oscillation at 160\,MHz,
yielding the amplitude $A(\theta,\varphi)$.}
\end{figure}

By probing $\nu$ at different times $t$, we find
$B_\textrm{tot}(t)$ to oscillate with $t$ with the frequency $f$.
Figure \ref{fig:fig2}(a) shows TRFR scans for two different $t$.
This data is obtained on sample 1 in the center of the four gates
that are aligned with the $x$- and $y$-axis. With
$V_a=V_b$\,=\,1~V, the electric field is oriented along $[100]$ at
the position of the laser focus. $B_\textrm{ext}$ is applied along
$[110]$ ($\theta = 90^\circ$) . The TRFR signal monitors the
coherent precession of the QW electron spins about the total field
$B_\textrm{tot}(t)$. Spins precess faster for $t$\,=\,-1.4\,ns
than for $t$\,=\,1.7\,ns. This variation of the precession
frequency follows the oscillation of $\boldsymbol{E}(t)$, as
becomes evident when plotting the fitted spin precession frequency
$\nu$ as a function of $t$, see symbols in Fig.~\ref{fig:fig2}(b).
The data points fit to a sinusoidal oscillation at frequency
$f$\,=\,160\,MHz (solid line). Except from a much weaker
contribution at $2f$~\cite{MeierNatPhys2007}, we do not observe
higher harmonics in $\nu(t)$, from which we conclude that the
linear dependence of the effective spin-orbit magnetic field on
$k$ in Eq.~\ref{eq:SIABIA} is valid. Specifically, terms cubic in
$k$ can be neglected. Such terms can be significant for the
zero-field SO splitting~\cite{PRL.90.076807} at the Fermi wave
vector. In our case however, the effective SO magnetic field is
determined by the in-plane drift wave vector $k$, which is much
smaller than the quantized wave number perpendicular to the QW,
and therefore negligible cubic contributions are expected.

The obtained $\nu(t)$ can be converted into $B_{\textrm{tot}}(t)$,
the amplitude of its oscillation being given by
$A(\theta,\varphi)$ according to Eq.~\ref{eq:RootExpanded}. By
measuring $A(\theta,\varphi)$ at varying $\theta$ and $\varphi$ in
the center of the four gates and comparing to Eq.~\ref{eq:AB}, we
determine $B_\textrm{SIA} = -4.2$~mT, $B_\textrm{BIA} = 21.6$~mT
for sample 1, and $B_\textrm{SIA} = -8.5$~mT, $B_\textrm{BIA} =
21.1$~mT for sample 2. These values were obtained for a
gate-modulation amplitude of 2~V, corresponding to $E_0 =
2900$~V/m~\cite{MeierNatPhys2007}.

In a geometrical interpretation, $A(\theta,\varphi)$ is the
projection of $\boldsymbol{B}_\textrm{SIA} +
\boldsymbol{B}_\textrm{BIA}$ onto $\boldsymbol{B}_\textrm{ext}$.
Because the SO fields depend linearly on $E_0$,
$A(\theta,\varphi)$ is proportional to the projection of the local
field $\boldsymbol{E}_0$ onto an axis defined by
$(x,y)=((B_\textrm{BIA}-B_\textrm{SIA})\sin\theta,(B_\textrm{BIA}+B_\textrm{SIA})\cos\theta)$.
At constant $\theta$, a spatial map of $A(\theta,\varphi)$
therefore directly images one component of $\boldsymbol{E}_0$.
Figure~\ref{fig:fig1}{(e) and (f)} show the calculated
$A(\theta,\varphi)$ using the measured values for $B_\textrm{BIA}$
and $B_\textrm{SIA}$ as a function of $\varphi$ for $\theta =
90^\circ$ and $45^\circ$ for sample 1. In the first case, $A$ is
proportional to the projection of $E$ onto the $x$-axis, with an
amplitude of $B_\textrm{BIA}-B_\textrm{SIA}$. For $\theta =
45^\circ$, the projection is onto $\varphi \approx 34^\circ$. For
vanishing SIA, the highest visibility would be exactly at
$\varphi=45^\circ$.

\begin{figure}
\includegraphics[width=80mm]{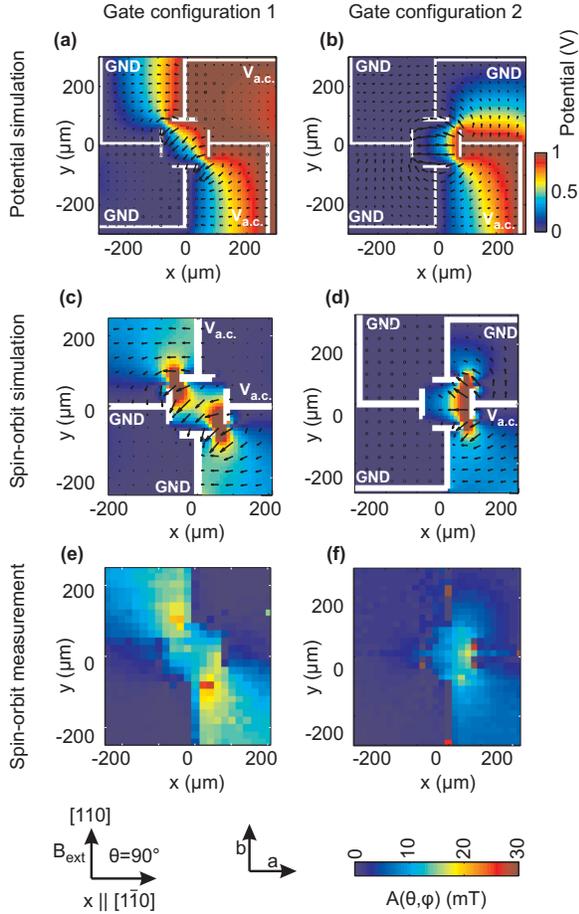}
\caption{\label{fig:fig3} (Color online) (a,b) Simulations of the
electric potential (color scale) and the electric field (arrows)
for two different gate potential configurations. (c,d) Simulated
effective SO field and electric field (arrows). (e,f) Measured
effective SO field on sample 1.}
\end{figure}

The electric field $\boldsymbol{E}$ is given by a superposition of
the electric fields induced by the two pairs of electrodes along
the $a$ and $b$ axis, and can be numerically determined with a
partial differential-equation solver (e.g. pdetool in Matlab) with
boundary conditions given by the voltages applied to the gate
electrodes and their connection lines. Figure~\ref{fig:fig3} shows
a situation for sample 1 with the gate electrodes oriented along
the $x$- and $y$-axis, and at two different bias configurations,
where either the top and the right gate electrode
[Fig.~\ref{fig:fig3}(a)] or only the right gate electrode
[Fig.~\ref{fig:fig3}(b)] were set to $V_0 = 1$~V, and all other
gates were grounded. In the center of the electrodes,
$\boldsymbol{E}_0$ points along the $[\overline{1}00]$-axis
($\varphi = 225^\circ$) or the $[\overline{1}10]$-axis ($\varphi =
180^\circ$), respectively, as expected from simply superposing two
uniform fields along the $x$ and $y$ axis that are proportional to
the two gate biases.

At every point in the two-dimensional sample plane, the local
electric field gives rise to a local $k$-vector, which allows the
calculation of $A(\theta,\varphi)$ using Eqs.~(\ref{eq:SIABIA})
and (\ref{eq:AB}). The result of this simulation is shown in
Fig.~\ref{fig:fig3}{(c) and (d)}. Considerable SO fields are also
expected away from the four gate electrodes, because of the
electric fields induced by the connection lines. The corresponding
measurements are shown in Fig.~\ref{fig:fig3}{(e) and (f)}. The
agreement with the simulation is good, except for the values of
$A$ measured close to a gate edge that are a factor of $\approx 2$
lower than in the simulation. There, the simulation assumes
perfect edges, leading to very high electric fields. In reality,
the edges are rough and round, and the electric field is lower.
Moreover, the simulations neglect the fact that the gates are
vertically offset from the QW by 30~nm, which compared to their
lateral separation of $150~\mu$m is, however, a negligible
distance.

\begin{figure}
\includegraphics[width=75mm]{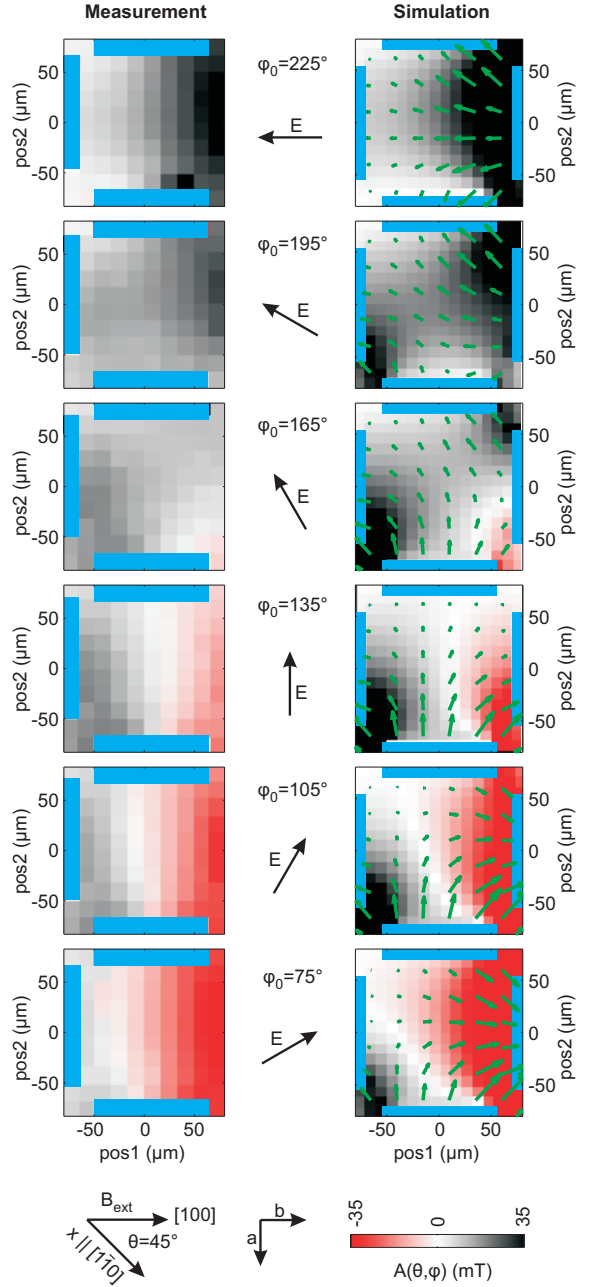}
\caption{\label{fig:fig4} (Color online) Measurement (left column)
and simulation (right column) of the SO effective magnetic field
between the four central gate electrodes (blue color) for sample
2. The left and the upper gate electrode are grounded, and biases
$V_a$ and $V_b$ are applied to the bottom and right electrode,
respectively. The direction of the electric field in the center of
the gate electrodes is indicated by black arrows. The calculated
electric field is visualized as green arrows (right column). The
magnetic field is applied along the $[100]$ axis, which points to
the right in the Figure. In the left column, the blue color is
used for positions where no TRFR signal could be observed because
the beams were blocked by the gate electrode. The color code is
different than in Fig.~\ref{fig:fig3} to visualize the sign of
$A$.}
\end{figure}

As expected from Fig.~\ref{fig:fig1}{(e)}, electric fields along
$x$ lead to high SO contributions to $\nu$. In
Figs.~\ref{fig:fig3}{(c) and (e)}, the connection lines along $y$
induce strong electric fields along $x$ in the upper left and the
lower right quadrant of the sample. These fields lead to a
$k$-vector in $x$-direction, and, as visible from
Fig.~\ref{fig:fig1}{(e)}, to a high $A$. In contrast, close to the
horizontal connection lines along $x$, the high electric fields in
$y$-direction do not lead to visible SO fields, as in this
situation, $\boldsymbol{B}_\textrm{SO}$ is perpendicular to
$\boldsymbol{B}_\textrm{ext}$ and therefore $A= 0$.

For the situation with three grounded gates
[Fig.~\ref{fig:fig3}{(f)}], outside of the four gates the electric
field component along $x$ is large in the full lower right corner
and partially in the upper right corner, leading to a high value
of $A$ at these positions. Inside the gates, $\boldsymbol{E}_0$ is
mainly aligned along $-x$ and decays from right to left because of
the shielding from the upper and lower gates which are at ground
potential, which is directly observable in the measured and
simulated maps of $A$.

In Fig.~\ref{fig:fig4}, we focus on the area between the four gate
electrodes. We choose a configuration in which the gates and
therefore the $(a,b$)-axis are oriented at $45^\circ$ to the
($x,y$)-axis. The measurements were taken on sample 2, and for the
simulations the obtained values for $B_\textrm{BIA}$ and
$B_\textrm{SIA}$ are used. We set $\theta = 45^\circ$, such that
$A$ is sensitive to the component of $\boldsymbol{E}_0$
approximately along $\boldsymbol{B_\textrm{ext}}$, i.e. along the
[100]-axis [see Fig.~\ref{fig:fig1}{(f)}]. The angle $\varphi_0$
of the electric field in the center of the electrodes has been
rotated by $30^\circ$ in each step by varying the amplitudes
$V_a=-V_0\cos(45^\circ+\varphi_0)$ and
$V_b=-V_0\sin(45^\circ+\varphi_0)$ of the two phase-locked
oscillators, connected to the bottom and the right gate electrode,
respectively, with $V_0$\,=\,1\,V. This direction is indicated by
an arrow between the measurement (left column) and the simulation
(right column). The simulation is again in good agreement with the
measurement, except close to edges, where the simulated electric
field and therefore also the SO fields are higher than observed.

In this configuration, $A$ is sensitive to the horizontal
component of $\boldsymbol{E}_0$, and therefore, the highest $A$
are measured close to the right gate electrode (connected to
$V_b$), provided that $V_b$ is large enough. There, $A$ is
positive for $\varphi = 225^\circ$ and $195^\circ$ (positive
$V_b$) and negative for $\varphi = 105^\circ$ and $75^\circ$
(negative $V_b$). As can be seen from the simulations in the right
column of Fig~\ref{fig:fig3}, the sign of $A$ correlates with the
component of $\boldsymbol{E}_0$ along $[100]$. For $\varphi =
135^\circ$, the left and right electrodes are grounded, and
$\boldsymbol{E}_0$ points along the vertical direction in the
center of the gates. On both the left and the right side of the
center, $\boldsymbol{E}_0$ turns sideways towards the respective
grounded lateral electrode, leading to positive and negative
values for $A$ on the two sides. Even though the bottom electrode
induces strong electric fields, only a small $A$ is observed close
to the electrode, since the field is mainly oriented along
$[010]$, i.e. perpendicular to $[100]$. For $\varphi=165^{\circ}$,
$\boldsymbol{E}_0$ has both positive and negative components along
$[100]$ close to the right gates, therefore leading to the
appearance of both signs of $A$, as can be seen both in the
simulation and in the measurement. In the corners of the square
defined by the four gate electrodes, high $A$ are measured as long
as the two neighboring electrodes are on different potentials.
There, the electric field is diagonal and therefore has
substantial components along $[100]$.

In conclusion, we have shown that the electric field induced by
biased gate electrodes modifies the electron spin precession in an
InGaAs QW through the presence of an SO effective magnetic field.
Spatially resolved optical measurements of the spin precession
frequency are in good agreement with numerically obtained spatial
maps. They confirm that the change in spin precession frequency is
a measure of the projection of the drift momentum and thus the
electric-field along an in-plane direction that is given by the
Rashba and Dresselhaus constants as well as the direction of
$\boldsymbol{B}_\textrm{ext}$. By adjusting the angle of
$\boldsymbol{B}_\textrm{ext}$, different components of the
electric field can be spatially mapped. Alternatively, if the
electric field is known, this technique might allow to measure
spatial variations of the Rashba and Dresselhaus
constants~\cite{Sherman2003, Liu2006}.

We acknowledge R. Allenspach, M. Duckheim, T. Ihn, R. Leturcq, D.
Loss and M. Witzig for helpful discussions. This work was
supported by the Swiss National Science Foundation (NCCR Nanoscale
Science).
%


\end{document}